\documentclass[11pt,a4paper]{article}
\usepackage[T1]{fontenc}
\usepackage[latin1]{inputenc}
\usepackage[english]{babel}
\usepackage{amsmath}
\usepackage{epsfig}
\usepackage{amssymb}

\newcommand{\be}{\begin{equation}}
\newcommand{\ee}{\end{equation}}

\newcommand{\bea}{\begin{eqnarray}}
\newcommand{\eea}{\end{eqnarray}}
\newcommand{\ba}{\begin{array}}
\newcommand{\ea}{\end{array}}
\newcommand{\beqa}{\begin{eqnarray}}
\newcommand{\eeqa}{\end{eqnarray}}

\newcommand{\lsim}{\lesssim}
\newcommand{\gsim}{\gtrsim}

\newcommand{\matr}{\left( \begin{array}}
\newcommand{\ematr}{\end{array} \right)}

\newcommand{\nue}{{\nu_{e}}}
\newcommand{\num}{{\nu_{\mu}}}
\newcommand{\nut}{{\nu_{\tau}}}

\begin{document}


\mbox{}\vspace*{-1cm}\hspace*{9cm}\makebox[7cm][r]{JyFL-HE 8/2003, NORDITA-2003-50HEP}
\medskip

\Large

\begin{center} {\bf Effects of sterile neutrinos on the ultrahigh-energy cosmic neutrino flux}

\bigskip

\normalsize
{P. Ker\"anen\footnote{keranen@nordita.dk}}

{\it Nordita, Copenhagen, Denmark}

{J. Maalampi\footnote{jukka.maalampi@phys.jyu.fi}}

{ \it Department of Physics, University of Jyv\"askyl\"a, Finland} and
{\it Helsinki Institute of Physics, Helsinki, Finland}

\smallskip
 {M. Myyryl\"ainen\footnote{minja.myyrylainen@phys.jyu.fi}} and
{J. Riittinen\footnote{janne.riittinen@phys.jyu.fi}}

{ \it Department of Physics, University of Jyv\"askyl\"a, Finland}
\\[15pt]

\bigskip

\normalsize

{\bf\normalsize \bf Abstract}

\end{center}

\normalsize

We investigate the effect of sterile neutrinos that are nearly degenerate 
with active ones on the flux of ultrahigh-energy cosmic ray neutrinos at earth. 
This offers a way to probe neutrino oscillations in the mass-squared range
$10^{-16}{\rm eV}^2\lesssim \delta m^2\lsim 10^{-11}$~eV$^2$ which maybe hard 
to detect by any other means. Taking into account the present experimental 
uncertainties of the active-active mixing angles and by allowing any values 
for the active-sterile mixing angles we find that the ratio of the electron 
and muon neutrino fluxes may change by -40\% to 70\% in comparison with the 
ratio in the absence of active-sterile mixing.\bigskip

\newpage

\normalsize

\section{Introduction}

Neutrino telescopes will measure the relative fluxes of
ultrahigh-energy neutrinos $\nu_e$, $\nu_{\mu}$, $\nu_{\tau}$
created in distant astrophysical objects like active galactic
nuclei (AGN) and gamma ray bursts (GRB). The fluxes measured at
earth are in general expected to differ from the initial
fluxes at source due to the oscillations of neutrinos during
their flight~\cite{learned}. A general assumption is that the
initial flux at source has the composition as 
 $\Phi^0_{\rm e}:\Phi^0_{\mu}:\Phi^0_{\tau}= 1:2:0$ 
(making no distinction between neutrinos
and antineutrinos) since neutrinos
are thought to be produced in the pion decays $\pi\to\mu
+\nu_{\mu}$ and in the subsequent muon decays $\mu\to {\rm
e}+\nu_e +\nu_{\mu}$. The measurement of the flavour composition 
of ultrahigh-energy neutrinos in terrestrial detectors 
will thus provide us with information
about the neutrino mixings~\cite{bento},~\cite{yasuda}. In particular, if the atmospheric
neutrino mixing is maximal, as the data suggest, the observed
ratio is expected to be approximately $\Phi_{\rm
e}:\Phi_{\mu}:\Phi_{\tau}= 1:1:1$ irrespective
to the value of the solar mixing angle~\cite{bento,yasuda}. 
(For recent works on the subject, see e.g. 
~\cite{Barenboim:2003jm,George,vadim}).

A sterile neutrino, if mixing with active neutrinos, might have
large effects on the fluxes of ultrahigh-energy neutrinos, easily detectable 
in the future neutrino telescopes
\cite{bento}. The existing data on solar and atmospheric neutrinos
combined with the recent results of the KamLAND reactor experiment~\cite{KamLAND}
clearly rule out active-sterile mixing as a dominant mixing mode
of neutrinos~\cite{tomppa}. Nevertheless, substantial sterile
components in the neutrino flavour states are still allowed~\cite{vissani,smirnov,valle}. On the other hand, the recent
precise determination of various cosmological parameters, by the WMAP~\cite{WMAP}, 
2dFGRS~\cite{2dF}, CBI~\cite{CBI}, and ACBAR~\cite{ACBAR} experiments, 
puts strong constraints on the number of
relativistic particle species present at Big Bang Nucleosynthesis
and on the amount that neutrinos contribute to the critical
density of the universe. A population of sterile neutrinos in the
early universe, created and maintained by oscillations, would
contradict these constraints unless the active-sterile mixing
and/or the relevant mass-squared differences are sufficiently
small~\cite{Barbieri}. Whether there already exists a conflict with the LSND
oscillation result~\cite{LSND} is still in dispute~\cite{pierce,hannestad}.

In this paper we shall study the effects of active-sterile
neutrino mixing on the fluxes of ultra-high energy neutrinos on their arrival at
earth. It should be emphasized that irrespective of the fact
that sterile neutrinos are possibly playing just a marginal
role in the solar and atmospheric neutrino oscillation phenomena
and of whether the LSND result is verified or not, sterile
neutrinos may well exist in such regions of the $(\sin^2
2\theta,\delta m^2)$ parameter space that are not probed in these
experiments at all. As long as the mass-squared differences
are in the range $\delta m^2\gsim 10^{-16}$ eV$^2$, active-sterile
neutrino mixings can have measurable effects on the UHECR neutrino
fluxes. On the other hand, if the mass-squared difference is
smaller than about $10^{-11}$ eV$^2$, the smallest difference one
can probe with solar neutrinos, the active-sterile mixing would have
remained unnoticed in all existing experiments. The UHECR neutrinos offer a 
way to probe neutrino oscillations in the mass-squared  range $10^{-16}$
eV$^2\lesssim \delta m^2\lesssim 10^{-11}$ eV$^2$ which may be hard to detect 
by any other means. Our objective is to study the effects of oscillations in this range.

\section{The influence of neutrino oscillations on neutrino fluxes}

In three flavour framework the flavour fields $\nue, \num$, and
$\nut$ and the mass eigenfields $\hat\nu_1, \hat\nu_2$ and  $\hat\nu_3$ are
related by a $3\times 3$ mixing matrix $U$ as
\be
\left(
\begin{array}{c}
  \nu_e \\
  \nu_{\mu} \\
  \nu_{\tau} \\
\end{array}
\right)
=U \left(
\begin{array}{c}
  \hat{\nu_1} \\
  \hat{\nu_2} \\
  \hat{\nu_3} \\
\ea
\right),
\ee
where
\be
U=
\left(
\begin{array}{ccc}
  U_{e1} & U_{e2} & U_{e3} \\
  U_{\mu1} & U_{\mu2} & U_{\mu3} \\
  U_{\tau1} & U_{\tau2} & U_{\tau3} \\
\ea
\right) =
\left(
\begin{array}{ccc}
  c_{12}c_{13} & s_{12}c_{13} & s_{13} \\
  -s_{12}c_{23}-c_{12}s_{23}s_{13} & c_{12}c_{23}-s_{12}s_{23}s_{13} & s_{23}c_{13} \\
  s_{12}s_{23}-c_{12}c_{23}s_{13} & -c_{12}s_{23}-s_{12}c_{23}s_{13} & c_{23}c_{13} \\
\ea \right), \label{genU}
\ee
and, neglecting the possible CP
violation, $c_{jk} = \cos{\theta_{jk}}$ and $s_{jk} =
\sin{\theta_{jk}}$. The solar neutrino data together with the
recent KamLAND reactor data~\cite{KamLAND} single out the so-called large mixing
angle solution, with mixing angle $\theta_{12}$ bounded into the
range $0.49<\theta_{12}< 0.67$~\cite{Holanda}. On the other hand,
observations of atmospheric neutrinos are consistent with maximal
mixing of the mass eigenfields $\nu_2$ and $\nu_3$, allowing for a
mixing between them in the range $0.63<\theta_{23}< 0.94$
\cite{GonGarz}. The third mixing angle, $\theta_{13}$, is bounded
to small values, $0\leq\theta_{13}\leq 0.1$, by the
non-observation of oscillations in CHOOZ~\cite{chooz} and Palo
Verde~\cite{paloverde} experiments. A one-parameter matrix that is
consistent with these limits is
\be
\left(
\begin{array}{ccc}
  c_{12} & s_{12} & 0 \\
  -s_{12}/\sqrt{2} & c_{12}/\sqrt{2} & 1/\sqrt{2} \\
  s_{12}/\sqrt{2} & -c_{12}/\sqrt{2} & 1/\sqrt{2} \\
\ea \right),\label{idU}
\ee
which corresponds to $\sin
\theta_{13}=0$ and maximal atmospheric mixing $\sin
2\theta_{23}=1$.

The observed flux of the neutrino flavour $\nu_{l}$ at earth, $\Phi_{l}$, is given by
\be
\Phi_{l}=\sum_{l'} P_{ll'}\cdot\Phi_{l'}^0 \,  , 
\ee where 
$P_{ll'}$ is the probability of the transition 
$\nu_{l'}\rightarrow \nu_{l}$,
\be
P_{ll'} =\delta_{l'l} - \sum_{i \neq j} U_{l'i}^{\ast} U_{li}
U_{l'j} U_{lj}^{\ast} (1-e^{-i {(m_{j}^2-m_{i}^2) L}/{2E}}) 
\ee
and $\Phi^0_{l}$ is the flux at source.
For UHECR neutrinos the distance of flight $L$ and
energy $E$ vary so that $P_{ll'}$ averages to
\be
P_{ll'}=\sum_{i} \vert U_{li}\vert^2 \vert U_{l'i}\vert^2 \, . 
\ee

If the mixing happens according to the matrix (\ref{idU}), the flux
ratio at the earth is $\Phi_{{\rm e}}:\Phi_{{\mu}}:\Phi_{{\tau}}= 1:1:1$ irrespective of
the value of the solar mixing angle $\theta_{12}$. When the values
of the mixing angles $\theta_{13}$ and $\theta_{23}$ vary in the
range allowed by experimental data, this ratio may have values
within the gray boomerang-shaped area of Fig.~1 (see below). From
observational point of view the fluxes $\Phi_{{\rm e}}$ and
$\Phi_{{\mu}}$ are more interesting than the flux of the tau
neutrino as they can be determined probably more precisely than
$\Phi_{{\tau}}$. The present oscillation data allows their
ratio to vary in the range $0.75\lesssim \Phi_{{\rm e}}:\Phi_{{\mu}}\lesssim 1.17$, 
where the lower end corresponds
to $\theta_{23} = 54^{\circ}$ and the upper end to 
$\theta_{23} = 36^{\circ}$, which are the largest and smallest value of the mixing 
angle $\theta_{23}$ that experiments allow.

\section{Effects of sterile neutrinos}

We shall now consider the situation where there exists a sterile
neutrino (neutrinos) that mixes with active neutrinos so that the
ensuing new mass eigenstate is nearly degenerate with another mass
state. As mentioned earlier, if the mass-squared difference
$\delta m^2$ of the two states is in the range $10^{-16}$
eV$^2\lesssim\delta m^2\lsim 10^{-11}$ eV$^2$, the existence of the
sterile flavour would have escaped detection in the oscillation
experiments performed so far, but they could, at least in
principle, have measurable effects on the fluxes of the UHECR
neutrinos. We will now study how large these effects could be.

Let us first assume that there exists just one sterile neutrino $\nu_s$,
which mixes with the superposition of the active states that constitutes the
mass eigenstate $\hat\nu_3$. Let us denote the mass eigenstates in this 
four-flavour scheme as follows:
\begin{eqnarray}
  \nu_1 &= &\hat{\nu_1} \nonumber \\
  \nu_2 &= &\hat{\nu_2} \nonumber \\
  \nu_3 &= &\cos{\varphi}\ \hat{\nu_3}- \sin{\varphi}\ \nu_s  \nonumber \\
  \nu_4 &= &\sin{\varphi}\ \hat{\nu_3} + \cos{\varphi}\ \nu_s \, .
\label{mstates}
\end{eqnarray}

We assume that the mass difference of the states $\nu_3$ and  $\nu_4$ is so 
small that in the processes, like particle decays, which are measured in 
laboratory experiments, these two states are not distinguished but appear as a 
single state with couplings equal to those of the state $\hat\nu_3$. If
$\vert \delta m_{34}^2\vert=\vert m_4^2-m_3^2\vert\lsim 10^{-11}$ eV$^2$, the 
existence of the fourth state had not been revealed in neutrino oscillation 
experiments performed so far. In this range of the mass-squared difference 
there is no limitations for the value of the mixing angle $\varphi$ from the 
existing data.

The neutrino mixing matrix is in this case given by
\be
U^{(4)} =
\left(%
\begin{array}{cccc}
   U_{e1} & U_{e2} & \cos{\varphi}\  U_{e3} & \sin{\varphi}\   U_{e3}\\
 U_{\mu1} & U_{\mu2} & \cos{\varphi}\  U_{\mu3} & \sin{\varphi}\ U_{\mu3} \\
  U_{\tau1} & U_{\tau2} & \cos{\varphi}\  U_{\tau3} & \sin{\varphi}\ U_{\tau3} \\
 0& 0 &  -\sin{\varphi}  & \cos{\varphi} \\
\end{array}%
\right),
\label{U4}
\ee
where $U_{li}$ are elements of the $3\times 3$ matrix (\ref{genU}). This 
active-sterile mixing suppresses the transition probability $P_{ll'}$ 
according to
\be
P_{ll'}=\vert U_{l1}\vert ^2 \vert U_{l'1}\vert ^2+\vert U_{l2}\vert ^2 \vert U_{l'2}
\vert ^2+S_3\cdot\vert U_{l3}\vert ^2 \vert U_{l'3}\vert ^2 \, ,
\label{Pster}\ee
where $l,l'=e, \mu, \tau$ and  we have denoted
$
S_3=\cos^4{\varphi} + \sin^4{\varphi}
$.

The suppression factor has a value within $\frac{1}{2}\leq S_3\leq 1$ 
depending on the value of the mixing angle $\varphi$. This generalizes
straightforwardly to the case where all three mass states are accompanied, as 
a result of active-sterile neutrino mixing, by a nearly degenerate new mass 
state. In this case the transition probability is given by

\be
P_{ll'}=S_1\cdot\vert U_{l1}\vert ^2 \vert U_{l'1}\vert ^2+ S_{2}\cdot\vert U_{l2}\vert ^2 \vert U_{l'2}
\vert ^2+ S_3\cdot\vert U_{l3}\vert ^2 \vert U_{l'3}\vert ^2 \, ,
\label{Pster3}\ee
where 
\be S_i=\cos^4{\varphi_i} + \sin^4{\varphi_i}
\ee 
($i=1,2,3$) and $\varphi_i$ is the mixing angle between the mass 
state $\nu_i$ and the accompanying sterile state $\nu_{si}$. In 
general the factors $S_i$ varies in the range 
$\frac{1}{2}\leq S_i\leq 1$ independently of each other.

Obviously, this is not the most general mixing scheme between three active 
and three sterile neutrinos. In  general the mixing of six flavours depends 
on 15 rotation angles while we describe the mixing in terms of  just six 
angles ($\theta_{ij},  \varphi_i;\ i,j=1,2,3$). Nevertheless, we consider 
this as an interesting special case, and we also believe that it provides us 
with a generic picture of the effects of active-sterile mixings on the UHECR 
neutrino fluxes, at least as far as the size of the effect is concerned. 

A theoretically relevant question is how to achieve a high degeneracy between 
the active and sterile states at the same time with a large mixing. In the 
simplest scheme there are two mass scales involved, one ($m$) that gives the 
absolute mass scale and another ($\mu\sim \sqrt{\delta m^2}$) that determines 
the mixing. The mixing angle in such a case would obey $\varphi\sim \mu/m$, 
and it is small except in the case where the absolute mass scale and the mass 
difference are of the same order of magnitude. Without pursuing the issue 
further and considering more sophisticated mixing schemes we just note that 
in this simple picture only the lightest of the three standard neutrino mass 
states could have a sterile partner with $\delta m^2\lsim 10^{-11}$ eV$^2$ as 
the overall  scale of the neutrino mass spectrum, in contrast with the mass 
differences, is not fixed by the present experimental data leaving the mass 
of the lightest neutrino free.

\section{Results}

Let us now look at the effect of the active-sterile mixing on the fluxes of 
the active neutrinos measured in terrestrial detectors. The ternary plot of 
Fig.~1  presents  the relative fluxes $F_l$  ($l=e,\mu,\tau$) of the active 
neutrinos $\nu_e$, $\nu_{\mu}$, $\nu_{\tau}$ at earth, where

\be
F_l=\frac{\Phi_l}{\Phi_e+\Phi_{\mu}+\Phi_{\tau}} \, .
\ee

It is assumed in this plot that the initial fluxes of different flavours at 
source are given by the standard ratio 
$\Phi^0_{\rm e}:\Phi^0_{\mu}:\Phi^0_{\tau}= 1:2:0$. The gray area corresponds 
to the situation where all the sterile neutrinos decouple, i.e. 
$\varphi_i =0$ for all $i=1,2,3$ (or the degeneracies are such that 
$\delta m^2\lsim 10^{-16}$ eV$^2$), and the mixing angles $\theta_{ij}$ of 
the $3\times 3$ mixing matrix are varied in their phenomenologically allowed 
ranges ($0.49<\theta_{12}< 0.67$, $0.63<\theta_{23}< 0.94$ and 
$0\leq\theta_{13}\leq 0.1$). (We shall call this the SM case.) The black area 
is obtained by allowing the active-sterile mixing angles $\varphi_i$ vary
freely (the sterile-mixing case).

\begin{figure}[t]
\centering
\epsfig{file=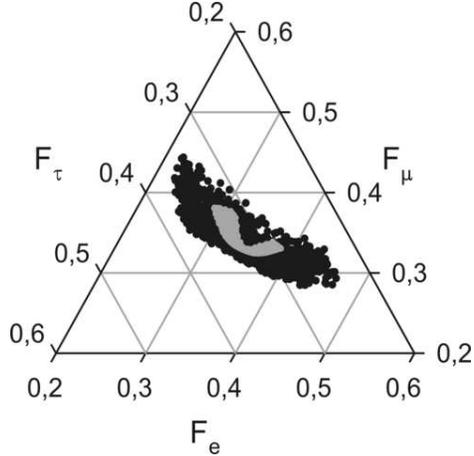,height=6cm}
\caption{Relative fluxes of the UHECR neutrinos at earth. The gray area 
corresponds to the situation with no sterile mixing and the standard mixing 
angles $\theta_{ij}$ varying within the experimentally allowed ranges. The 
black area, in part behind the gray area, corresponds to the situation with 
three sterile neutrinos with arbitrary mixing angles $\varphi_i$.}
\label{fig1}
\end{figure}

%

The phenomenologically most interesting is the flux ratio of the electron and 
muon neutrinos,
\be
R_{e\mu}\equiv\frac{\Phi_{\rm e}}{\Phi_{\mu}} \, .
\ee
In the SM case this ratio varies in the 
range 
\be
0.75\lsim (R_{e\mu})_{\rm SM}\lsim 1.17 \, ,
\label{RSM}\ee
for the present accuracy of the mixing angles $\theta_{ij}$. In the 
sterile-mixing case one has 
\be
0.47\lsim (R_{e\mu})_{\rm sterile}\lsim 1.62 \, .
\label{Rster}
\ee
The relative deviation of the ratio $R_{e\mu}$ from its SM value, caused by 
the sterile mixing, varies approximately in the range 
\be
-40\%\lsim \frac{(R_{e\mu})_{\rm sterile}-(R_{e\mu})_{\rm SM}}{(R_{e\mu})_{\rm SM}}\lsim 70\% \, .
\label{deviation}
\ee

The largest positive deviation is achieved when the sterile mixing angle 
$\varphi_1$ is very small and the other two $\varphi_2$ an $\varphi_3$ are 
nearly maximal (a representative set of  values being $\varphi_1=3^0,\ 
\varphi_2=37^0,\ \varphi_3=44^0$), while the largest negative deviation is 
obtained when $\varphi_1$ and $\varphi_2$ are nearly maximal and $\varphi_3$ 
is very small ($\varphi_1=43^0,\ \varphi_2=45^0,\ \varphi_3=0.02^0$). In the 
former limit the flux ratios are approximatively 
$\Phi_{\rm e}:\Phi_{\mu}:\Phi_{\tau}= 0.4:0.3:0.3$, in the latter limit 
$\Phi_{\rm e}:\Phi_{\mu}:\Phi_{\tau}= 0.2:0.5:0.3$. The effect of the 
active-sterile mixing to the relative fluxes can thus be quite large and 
easily detectable in the future neutrino telescope experiments.

Let us consider the situation of just one sterile neutrino mixing with active 
neutrinos with the other two sterile neutrinos decoupling. Such a situation 
is described by the mixing matrix~(\ref{U4}), where $\varphi_1=\varphi_2=0$.  
In Fig.~2 we present a ternary plot corresponding to this situation. To
see how the deviation from the SM case depends on the active-sterile mixing 
angle, we have plotted in Fig.~3 the maximum and minimum values of the ratio 
$R_{e\mu}$ as a function of $\varphi_3$ when the rotation angles of the 
$3\times 3$ mixing matrix vary in their experimentally allowed ranges. As can 
be seen from this figure, the SM result and the sterile mixing result do not 
overlap when the active-sterile mixing is close to maximal. The flux of 
electron neutrinos increases in comparison with  muon neutrinos with 
increasing $\phi_3$. At the same time the relative flux of tau neutrinos 
remains nearly unchanged.

\begin{figure}[t]
\centering
\epsfig{file=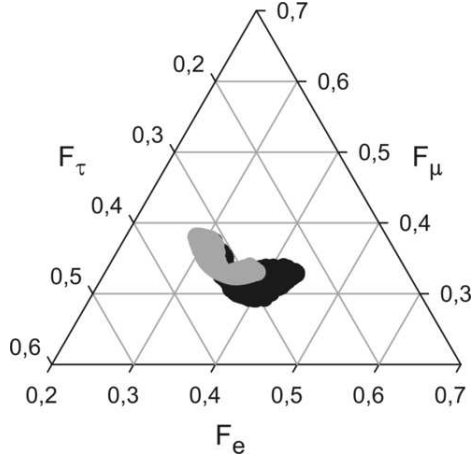,height=6cm}
\caption{Relative fluxes of the UHECR neutrinos at earth. The gray area 
corresponds to the situation with no sterile mixing and the standard mixing 
angles $\theta_{ij}$ varying within the experimentally allowed ranges. The 
black area corresponds to the situation with one sterile neutrino with 
arbitrary mixing  with the heaviest neutrino.}
\label{fig2}
\end{figure}

%

%
\begin{figure}[t]
\centering
\epsfig{file=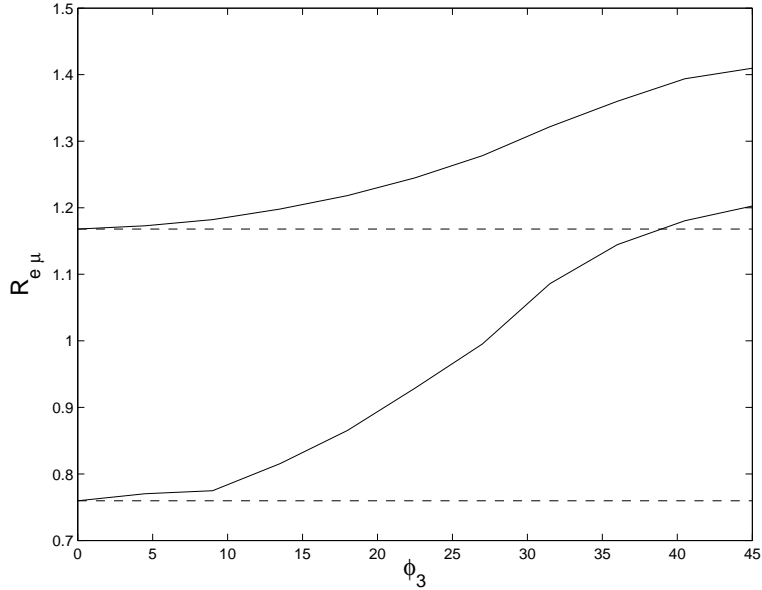,height=8cm}
\caption{The ratio of the fluxes of electron and muon neutrinos as a function 
of the sterile mixing angle $\varphi_3$. The area between the solid lines 
corresponds to the experimental uncertainty of the standard mixing angles 
$\theta_{ij}$. The dashed lines indicate the situation in the case of no 
sterile mixing.}
\label{fig3}
\end{figure}

%

In the near future there will be many new experiments 
that will improve our knowledge of the values of the mixing angles 
$\theta_{ij}$. Let us explore how this would affect the sensitivity of UHECR 
neutrino measurements in probing the sterile mixing. We 
follow~\cite{Barenboim:2003jm} and anticipate the following ranges for the mixing
angles: $0.54<\theta_{12}<0.63$,\, 
$\frac{\pi}{4}\times0.9<\theta_{23}<\frac{\pi}{4}\times1.1$ 
and $0<\theta_{13} <0.1$. The relative 
fluxes of neutrino flavours at earth in this case are presented in Fig.~4, 
where the gray area corresponds to the situation with no active-sterile
oscillations. The ratio of the electron neutrino and muon neutrino fluxes 
varies in the absence of sterile mixing in the range 
$0.88\lsim (R_{e\mu})_{\rm SM}\lsim 1.13$, while in the presence of three
sterile-neutrino oscillations with arbitrary mixing angles the variation 
happens within the range $0.58\lsim R_{e\mu}\lsim 1.57\,$. Comparison with 
(\ref{RSM}) and (\ref{Rster}) reveals that improving the accuracy of the 
determination of $\theta_{ij}$ would not substantially affect the 
distinctiveness of  sterile neutrino signal in the UHECR neutrino flux.

\begin{figure}[t]
\centering
\epsfig{file=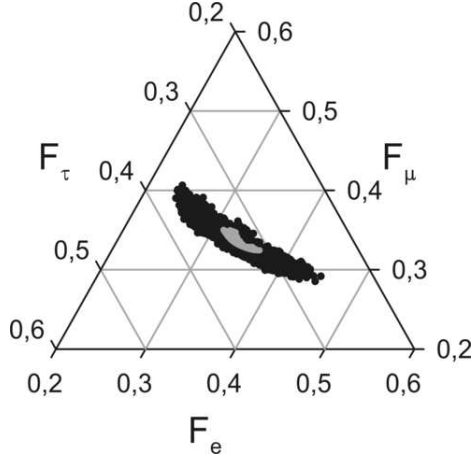,height=6cm}
\caption{Relative fluxes of the UHECR neutrinos at earth. The gray area 
corresponds to the situation with no sterile mixing and the standard mixing 
angles $\theta_{ij}$ varying within the anticipated sensitivity ranges of 
future experiments. The black area corresponds to the situation with three 
sterile neutrinos with arbitrary mixing angles $\varphi_i$.}
\label{fig4}
\end{figure}

%

\section{Discussion}

If one assumes the conventional production mechanism of ultrahigh-energy 
neutrinos in astrophysical sources, the three neutrino flavours $\nu_e$, 
$\nu_{\mu}$, $\nu_{\tau}$  arrive at earth  approximately in the flux ratios 
1:1:1. We have investigated how these ratios are changed if the active 
flavours form with hypothetical sterile neutrinos nearly degenerate mass 
state pairs with mass-squares in the range 
$10^{-16}$ eV$^2\lsim\delta m^2\lsim 10^{-11}$ eV$^2$. 
The effect would be detectable in future experiments, such as the IceCube~\cite{IceCube}, 
ANTARES~\cite{ANTARES} and NESTOR~\cite{NESTOR}, which will improve the sensitivity to ultrahigh-energy 
neutrinos by 2 to 3 orders of magnitude during the next decade~\cite{Spiering}. 

We have tacitly assumed in our analysis that neutrinos do not decay during 
their flight from the source to terrestrial detectors. In the standard model 
framework this certainly is a safe assumption as the experimental limits on 
the magnetic moments of neutrinos imply that the decays $\nu_i\to \nu_j + 
\gamma$ are very slow and completely irrelevant for our flux considerations. 
If one extends the model to include majorons ($J$), light scalar particles 
that appear in theories where  global lepton symmetry in spontaneously
broken~\cite{gelmini}, new decay channels, $\nu_i\to \nu_j + J$, are open 
that could be rapid enough to cause detectable modulation of neutrino fluxes 
\cite{keranen}. If these decay do take place their effect could overrule the 
effect of the sterile mixing considered in this paper~\cite{beacomA}.

We have assumed above that the initial flux composition is the canonical one, 
i.e. $\Phi^0_{\rm e}:\Phi^0_{\mu}:\Phi^0_{\tau}= 1:2:0$. It has been argued 
that the muons produced in pion decay can lose energy e.g. in strong magnetic 
field of the source so that no ultrahigh-energy electron neutrinos are 
available~\cite{Rachen} (see also~\cite{Horvat}). In this situation the 
initial flux composition is $\Phi^0_{\rm e}:\Phi^0_{\mu}:\Phi^0_{\tau}= 
0:1:0$. We have checked this case, and it turned  out that there the effect 
of sterile mixing is smaller than in the canonical case: in the presence 
(absence) of sterile neutrinos one has $0.19\lsim R_{e\mu}\lsim 1.10$ 
($0.25\lsim R_{e\mu}\lsim 0.93$).

Let us finally consider the existence of the three sterile neutrinos of our 
picture from the point of view of cosmology. One possible cosmological effect 
of sterile neutrinos is related to the Big Bang nucleosynthesis. As is well 
known~\cite{Barbieri} active-sterile mixing might bring - depending on the 
values of the mixing angle and mass-squared difference - sterile neutrinos in 
equilibrium with active neutrinos before neutrino decoupling and the 
resulting excess in energy density would endanger the standard scheme for
the nucleosynthesis of light elements. This argument leads to the following 
bound for $\nu_e \leftrightarrow \nu_s$ mixing~\cite{Shi}:
$|\delta m^2| \sin^2 2\alpha <5 \times 10^{-8}\ \mbox{eV}^2$
and to slightly less stringent bounds for other flavours. Obviously, this 
constraint has no consequence in our case as we assume 
$\delta m^2 < 10^{-11}$ eV$^2$. In other words, neutrino oscillations will 
not produce and maintain a significant sterile neutrino population. This also 
means that the conclusions one draws from the first results from WMAP 
satellite experiment~\cite{WMAP}, are not affected by the sterile neutrinos 
we are considering. The WMAP experiment, which measures the fluctuations of 
the cosmic microwave background radiation, leads to the bound 
$\Omega_{\nu}h^2=\sum_i m_i/93.5 {\rm eV}< 0.0076$ (95 \% CL) for the total 
energy density of neutrinos. This implies the limit $\sum_i m_i < 0.69$ eV 
for the sum of masses of all neutrino species. In our case only half of six 
neutrino species contribute to this sum as the number density of sterile 
neutrinos is negligible. Hence, when the bound is saturated, all neutrinos 
are nearly degenerate and all three masses $m_i$ ($i=1,2,3$) have the upper 
limit $m_i < 0.23$ eV, just like in the case of no sterile neutrinos. 

\section{Acknowledgements} We are grateful to Kimmo Kainulainen, University 
of Jyväskylä, for useful discussions. The work has been supported by the 
Academy of Finland under the contract no.\ 40677.

\end{document}